\documentclass[
 reprint,
 amsmath,amssymb,
 aps,
prx,
]{revtex4-2}

\usepackage{graphicx}
\usepackage{dcolumn}
\usepackage{bm}
\begin{document}

\preprint{APS/123-QED}

\title{Equilibrium Spin Polarization Arising From Chirality}

\author{Pius M. Theiler}
\email{pius.theiler@nrel.gov}
\affiliation{%
National Renewable Energy Laboratory,
Golden, CO, USA
}%

\author{Matthew C. Beard}

\affiliation{%
National Renewable Energy Laboratory,
Golden, CO, USA
}%

\date{28. October 2025}
\begin{abstract}
Chirality-induced spin selectivity (CISS) describes how chiral molecules and materials generate spin polarization even at thermal equilibrium. This observation has challenged established principles of microscopic reversibility and Onsager reciprocity. 
We resolve this paradox by formulating a pseudo-Hermitian quantum framework in which structural chirality and electron correlations are sufficient to produce CISS observables.  
Chirality enters through a non-local metric~$\eta$ that couples spin and spatial motion, leading to real spectra, unitary evolution, and thermodynamic consistency.  
The framework predicts a chirality-induced spin magnetic ordering characterized by a spin--displacement order $\langle \sigma \cdot x \rangle$, which reconciles equilibrium spin polarization with detailed balance and explains the persistence of CISS in materials composed of light elements. We also derive generalized Onsager-Casimir relations that respect the observed parity ($\mathcal{P}$) and time-reversal ($\mathcal{T}$) breaking, while preserving combined $\mathcal{PT}$-symmetry. This approach establishes a coherent foundation for equilibrium CISS and provides a route to link chemical chirality with measurable spin-to-charge conversion effects.
\end{abstract}

\maketitle

\section{Introduction}

Chirality‑induced spin selectivity (CISS) refers to the coupling between a system’s structural chirality and the spin orientation of electrons \cite{Aiello2022,Evers2022,Bloom2024,Foo2025}.
Originally discovered in non-equilibrium settings such as photoemission\cite{Ray1999} and transport experiments\cite{Xie2011}, CISS has later  also been reported under conditions that are nominally at thermal equilibrium\cite{BenDor2017,Ghosh2020,Theiler2023} and are distinct from electrical magnetochiral anisotropy\cite{Rikken2023}. Examples include persistent spin polarization manifested as induced magnetization \cite{BenDor2017,Zhu2024}, contact–potential shifts\cite{Abendroth2019,Ghosh2020,Theiler2023} and quantum capacitance\cite{Theiler2023}, and zero-bias magnetoresistance \cite{Kulkarni2020,Al-Bustami2022,Safari2024,haque2025,Tirion2024,Malatong2023}. This questions the hypothesis that CISS is only induced under non-equilibrium, dynamic conditions and disappears under or near equilibrium conditions. Under the dynamic perspective, many of these observations appear paradoxical. Textbook thermodynamics identifies equilibrium with the absence of entropy production together with the concept of microscopic reversibility and detailed balance\cite{Callen1985,Kubo1992}, which is often wrongly equated with time–reversal (\(\mathcal{T}\)) invariance of the Hamiltonian and Hermicity.

The experimental observation of $\mathcal{T}$-breaking  has fueled a lively debate in the CISS community and is tied up in the ongoing debate about the mechanism of CISS.  Discussions\cite{Yang2021,Hedegrd2023,Fransson2025,Sarkar2025,Aharony2025} often center around the Casimir-Onsager reciprocity relations\cite{Onsager1931,Onsager1931a,Casimir1945} and of Bardarson’s theorem\cite{Bardarson2008}.  The former provides a relation between flows and forces in thermodynamic systems out of equilibrium but where a notion of local equilibrium exist and linear response is valid\cite{Onsager1931,Onsager1931a,Casimir1945}. Recently, they were generalized to quantum spin systems that break-time reversal symmetry\cite{Bonella2014,DeGregorio2017,Luo2020,Huang2025}. The latter proves that in a two-terminal, elastic, time-reversal-symmetric conductor the transmission eigenvalues occur in degenerate Kramers pairs, seemingly forbidding net spin filtering in closed shell systems\cite{Bardarson2008}. As a result, the experimental evidence\cite{BenDor2017,Ghosh2020,Theiler2023,Zhu2024,Abendroth2019,Rikken2023,Ghosh2020,Kulkarni2020,Al-Bustami2022,Safari2024,haque2025,Tirion2024,Malatong2023} for spin polarization is difficult to reconcile with such constraints.  Here we address the question of how can equilibrium chirality-induced spin magnetic order exist without violating the second law of thermodynamics?

We think that much of the controversy originates in a subtle but important misconception:  
microscopic reversibility, the condition required for thermodynamic equilibrium, is not identical to invariance of the Hamiltonian under the anti-unitary \(\mathcal{T}\) operator. Wigner\cite{Wigner1931} demands only a unitary and an anti-unitary operation constrained by the symmetry of the system\cite{Dresselhaus2008}, which admits $\mathcal{T}$-breaking, but nontheless leads to stable quantum phases like superconductivity\cite{JBardeen1957}, ferromagnetism\cite{Dyson1956} or alter-magnetism\cite{Liu2022}. Some authors simplify the analysis by assuming that time-reversal symmetry and microscopic reversibility are synonyms\cite{Pathria1972,Huang1987,Pathria2011,Schwabl2008,Luo2020}. However, equilibrium requires detailed balance or microscopic reversibility and the stationarity of the Gibbs state with maximized entropy\cite{SakuraiQM,Lifshitz1981,Guenault1995}, none of which require that \( [H,\mathcal{T}]=0 \) or Hermicity\cite{Gardas2016}. 

\begin{figure}
\centering
\includegraphics[width=\columnwidth]{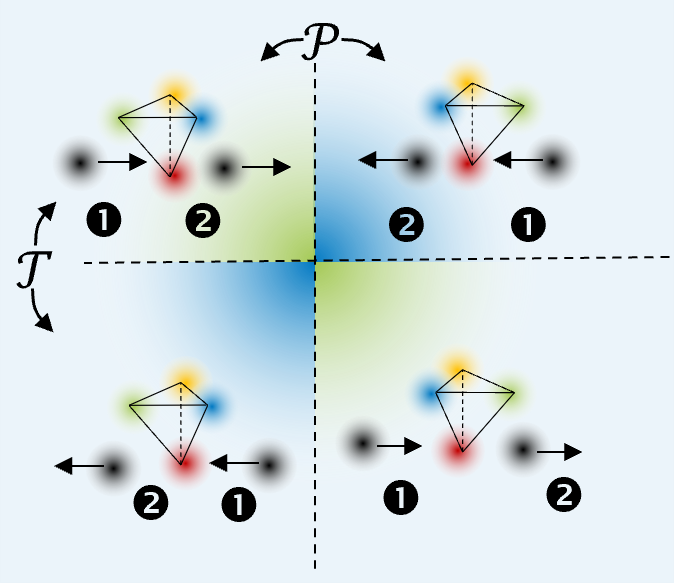}
\caption{Scattering from a chiral object. A scattering process involving a chiral material. A particle traveling form  $1 \rightarrow 2$ (top left) scatters from a chiral object (here denoted as a thetrahedron whose four vertices are distinct). Microscopically reversing that process, (bottom left), i.e., applying the time-reversal operator $\mathcal{T}$ produces a distinct sequence of events. To fully reverse the process requires a combined parity, $\mathcal{P}$ and $\mathcal{T}$, or $\mathcal{PT}$ operation. Only in the $\mathcal{PT}$ case the sequence of events (scattering off from the green before blue) is reversed correctly.  However, the in chiral systems the $\mathcal{P}$ operation transforms the system from one enatiomer to the other.  This paradox illustrates the complexity of applying microscopic reversibility in a chiral system. Arrows show the direction of movement of the point-like particle.  }
\label{Fig:PT-scattering}
\end{figure}

Indeed, systems that individually break spatial inversion (\(\mathcal{P}\)) and time reversal ($\mathcal{T}$) may respect the combined \(\mathcal{PT}\) operation, which is also an antiunitary operator and it is the composite symmetry that guarantees microscopic reversibility.  Lifshitz and Pitaevskii \cite{Lifshitz1981} and Sakurai\cite{SakuraiQM} long ago identified  an apparent contradiction between microscopic reversibility, detailed balance, and time-reversal in chiral systems illustrated in  Fig. \ref{Fig:PT-scattering}  which depicts a scattering event with a chiral tetrahedron. One can easily see that an electron traveling from left-to-right (1-2) in the top left panel sees a different sequence in potential then when the electron travels from right-to-left under time-reversal symmetry (bottom left panel), where the motion of the particles are reversed but their orientation preserved.  Recovery of the correct symmetry requires a combined $\mathcal{PT}$ symmetry (compare the top-left panel to the bottom-right panel).  However, these authors correctly point out that in a chiral system of just one enantiomer the $\mathcal{P}$ operation maps the system to the other enantiomer generating a completely different system.  Thus, the concept of microscopic reversibility in the strictest sense breaks down even when $\mathcal{PT}$-symmetry holds, and in these cases detailed balance conditions should be carefully considered. The situation is different when considering spin. Spin is a spinor and space is a polar vector transforming differently under $\mathcal{P}$, and $\mathcal{T}$ operations\cite{Liu2022}. 

Barron analyzed detailed balance conditions in the context of absolute asymmetric synthesis~\cite{Barron1987}: He showed that when prochiral molecules react in an initially isotropic system, the resulting chiral products form a racemic mixture without enantiomeric excess. In this case, the full system of reactants and products preserves $\mathcal{PT}$ symmetry, since $\mathcal{P}$ maps one enantiomer to the other and Barron derived detailed balance conditions under this $\mathcal{PT}$ symmetry. Barron argued that a  'true' chiral influence is required to break the symmetry in order to achieve absolute asymmetric synthesis. However, he specifically did not consider processes occuring in already chiral multi-electron systems with spin, where $\mathcal{P}$, $\mathcal{T}$, and $\mathcal{PT}$ are all broken, and it is precisely this situation we address here. 

To find detailed balance conditions in structurally chiral systems, we note that equilibrium is ensured whenever the Gibbs ensemble is invariant under the symmetry preserving the dynamical probabilities, even if that symmetry is neither $\mathcal{T}$ nor $\mathcal{PT}$. This is analogous to ferromagnetism~\cite{Dyson1956} or superconductivity~\cite{JBardeen1957}, where the equilibrium state breaks $\mathcal{T}$ but still minimizes the free energy and produces no entropy, resulting in degenerate Gibbs states related by $\mathcal{T}$~\cite{Callen1985} (e.g., "up" and "down" magnetization). Similarly, altermagnetism~\cite{Smejkal2022} demonstrates that composed symmetries can mix spin and spatial symmetries even without spin-orbit coupling~\cite{Liu2022}. In CISS, the relevant degenerate Gibbs states are separated by $\mathcal{P}$ and $\mathcal{T}$ breaking, that should allow for a description of spin selectivity at equilibrium\cite{Theiler2025}, which we term \emph{chirality-induced spin magnetism} (cismagnetism).

Previous approaches capture many aspects of CISS but often rely on additional assumptions combining environmental coupling~\cite{Dalum2019,Varshney2024,Volosniev2021,Zllner2020,Hedegrd2023,Savi2025,Sarkar2025,Aharony2025,Dalum2019,Hedegrd2023,Chiesa2024}, non-unitary dynamics~\cite{Matityahu2016,Utsumi2020,Zhao2025}, or correlated interactions~\cite{Fransson2019,Chiesa2024,Xu2024,Savi2025,Chiesa2025,Fransson2021,Fransson2025}. Our framework demonstrates that structural chirality (absence of mirror symmetry) combined with electron correlations alone suffices to reproduce both static magnetic order and the dynamic spin–charge phenomena observed experimentally, highlighting that CISS inherently goes beyond a single-particle picture\cite{Bloom2024,Foo2025}. Using a non-Hermitian Hamiltonian~\cite{Theiler2025} and Dyson mapping~\cite{Dyson1956,Jones-Smith2013}, we show that chirality can stabilize a static, $\mathcal{P}$- and $\mathcal{T}$-broken cismagnetic phase without violating thermodynamic principles and derive modified Onsager reciprocity relations for CISS.

\begin{figure*}
\centering
\includegraphics[width=0.80\textwidth]{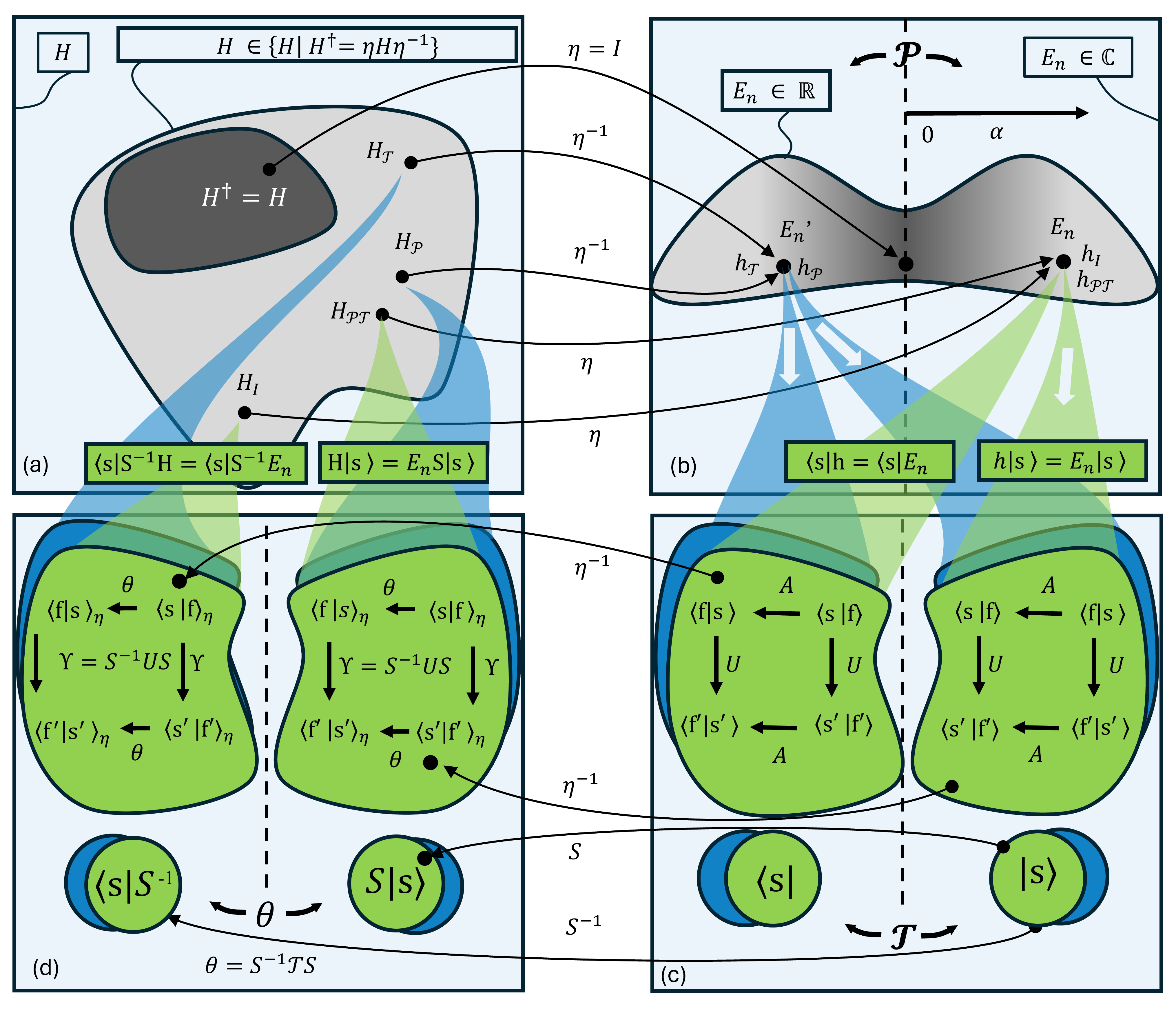}
\caption{Dyson mapping for structurally chiral Hamiltonians.  
(a) The physical space of all Hamiltonians $H$ includes pseudo-Hermitian cases ($H_I$, $H_T$, $H_P$, $H_{PT}$), with Hermitian Hamiltonians ($\alpha=0)$ forming a subset thereof.  
(b) In the mapped space, Hamiltonians are categorized according to their eigenenergies. Real eigenenergies correspond to the Hermitian and pseudo-Hermitian Hamiltonians and are represented by the shaded region. All chiral pseudo-Hermitian Hamiltonians can be parameterized by the order parameter $\alpha$, where $\alpha = 0$ corresponds to physical Hermitian systems.  
(c) For each eigenvalue, the corresponding left- and right-sided eigenvectors of the Dyson-mapped Hermitian Hamiltonians $h$ are obtained using the same unitary and antiunitary operations for both enantiomers. Green shading corresponds to the $\eta$-metric, blue shading to $\eta^{-1}$ metric sectors with equivalent mapping.  
(d) The physical Hilbert-space of the inverse Dyson mapping yields the physical eigenvectors of the pseudo-Hermitian Hamiltonians $H_I$ and $H_{PT}$ for green shading, which describe distinct physical systems but share an equivalent mapping structure. Blue shading would correspond to $H_{P}$ and $H_{T}$. The Dyson mapped unitary and antiunitary operators are defined within each sector (i.e., they do not allow crossing to another sector) and allow for defining the microscopic reversibility symmetry and detailed balance within each sector, resolving the paradox presented in Fig. \ref{Fig:PT-scattering}.}
\label{Fig:Classification}
\end{figure*}

\section{Model Hamiltonian}

Quantum systems often defy classical intuition through their intrinsic non-locality, demanding approaches\cite{Jones-Smith2013,Michishita2020} beyond heuristic reasoning. In Heisenberg’s spirit~\cite{Heisenberg1925}, when existing tools fail, one must have the courage to reconsider assumptions. Here we try to do so by reexamining the link between chirality and spin to uncover new insights into CISS. The internal symmetries of solid state systems provide a framework to understand their electronic, spin, and topological properties\cite{Kishine2022}. Figure~\ref{Fig:Classification}a illustrates the hierarchical inclusion of Hamiltonian classes which are grouped into three broad groups: (1) Hermitian Hamiltonians (dark grey, Fig. \ref{Fig:Classification}a) classified via symmetry under the Altland--Zirnbauer (AZ) classification~\cite{Altland1997} with standard inner products, which evolve unitarily, conserve energy, and underpin mean-field and quasi-particle formalisms.(2) The non-Hermitian extension by Kawabata \emph{et al.}~\cite{Kawabata2019} generalizes the symmetry classification to 38 classes for globally defined Bloch Hamiltonians, including the ten Hermitian AZ classes as limiting cases, encompassing pseudo-Hermitian systems with real spectra but correlations that go beyond quasi-particle formalism\cite{Jones-Smith2013,Michishita2020} (Fig. \ref{Fig:Classification}a light grey area)  and (3) genuinely dissipative systems with complex eigenvalues (Fig. \ref{Fig:Classification}a outside light-blue area) that allow for energy flow from and to their environment.  
  
A Hamiltonian $H$ is called a pseudo-Hermitian\cite{Mostafazadeh2002,Mostafazadeh2002a,Kawabata2019,Montag2024} if there exists a positive-definite Hermitian metric operator $\eta$ such that
\begin{equation}
    H^\dagger = \eta H \eta^{-1}, \qquad \eta=\eta^\dagger>0,
\end{equation}
and allow for a complete biorthogonal system of eigenstates. 
Equivalently, one can introduce a similarity transformation $S$ such that
\begin{equation}
    h = S H S^{-1}, \qquad \eta = S^\dagger S,
    \label{Eq. h transform}
\end{equation}
with $h$ now being Hermitian and with the identical (real) eigenenergies of $H$. This transformation, $H \rightarrow h$, is called Dyson-mapping\cite{Dyson1956,Jones-Smith2013} (shown schematically in Fig. \ref{Fig:Classification}a to \ref{Fig:Classification}b).  In this framework $\eta$ defines the physical inner product\cite{Mostafazadeh2002}
\begin{equation}
    \langle \psi | \phi\rangle_\eta := \langle \psi | \eta | \phi \rangle,
\end{equation}
and guarantees unitary time evolution (Fig. \ref{Fig:Classification} d).

Importantly, pseudo-Hermiticity\cite{Mostafazadeh2002} is sufficient to ensure a consistent thermodynamic description. Gardas \emph{et al.} ~\cite{Gardas2016} have proven that equilibrium as well as non-equilibrium identities of quantum thermodynamics hold without modification for quantum systems described by pseudo-Hermitian Hamiltonians, and that the Carnot statement of the second law of thermodynamics remains valid, provided the spectrum of $H$ is real and the metric $\eta$ is positive-definite so that the two-time work definition applies unchanged. This is a powerful insight, since it allows us to investigate a Hamiltonian, map it using the Dyson transform (Fig.\ref{Fig:Classification} from a to b), study the eigenfunctions under principles of Hermitian systems and then perform an inverse Dyson mapping to investigate  behavior of structurally chiral systems (Fig. \ref{Fig:Classification}c to  \ref{Fig:Classification}d).

We begin our investigation of chiral systems with a Hermitian Hamiltonian describing a multi-electron system. Without loss of generality, this discussion can be formulated for a closed-shell system with $2n$ electrons. Here, for simplicity, we focus on the two-electron case, 

\begin{equation}
H = 
\sum_{i=1}^{2}
\left[
\frac{p_i^2}{2m}
+ V_H(x_i)
\right]
+  V(x_1,x_2).
\end{equation}
 \(V_H(x_i)\) denotes the Hartree potential acting on each electron, and the last term accounts for the Coulomb repulsion between them.

Following theoretical works suggesting intrinsic electronic correlations in chiral systems~\cite{Fransson2019,Chiesa2024,Xu2024,Savi2025,Chiesa2025,Theiler2025}, we adopt a correlation ansatz~\cite{Jones-Smith2013}.  We assume that  spin and momentum couples without invoking conventional spin–orbit coupling through \(\sigma_1\!\cdot\!p_1=-\sigma_2\!\cdot\!p_2\). Under suitable conditions~\cite{Michishita2020}, the resulting non-Hermitian formulation is equivalent to a self-energy–corrected Green’s function, where correlations are encoded in the complex self-energies\cite{Fransson2025}. The correlation can be incorporated into the Hamiltonian through a productive zero,

\begin{equation}
H = 
\sum_{i=1}^{2}
\left[
\frac{p_i^2}{2m} 
+ i\,\alpha\, \sigma_i \!\cdot\! p_i
+ V_H(x_i)
\right]
+ V(x_1,x_2).
\end{equation}
Focusing on a single electron and incorporating the Coulomb interaction into an effective mean-field potential, the system reduces to a non-Hermitian single-particle Hamiltonian. This approach captures the essential effects of electronic correlations, which are generally difficult to treat exactly, and is consistent with the effective non-Hermitian descriptions emerging in strongly correlated systems~\cite{Michishita2020, Theiler2025}:

\begin{equation}
    H = \frac{p^2}{2m} + i\alpha\, \sigma\!\cdot\! p  +V(x),
    \label{eq:Hamiltonian}
\end{equation}
where $p=-i\hbar \partial_x$ is the momentum operator, $m$ the electron effective mass, $\sigma$ the Pauli matrices with the dot product selecting the component along the momentum direction, and $\alpha$ a chirality-induced spin drift velocity quantifying the strength of the spin–chirality coupling. While $\alpha$ is related to $G_0$, the magnitude of the electric toroidal monopole of the potential $V(x)$~\cite{Kishine2022,Inda2024,Kusunose2024}, the precise functional dependence relating $\alpha$ to $G_{0}$ remains unknown. For chiral systems, $V(x)$ is a real-valued, structurally chiral potential absent of any mirror-plane. If there is a mirror plane within the potential, then $G_0=0$ and concurrently $\alpha=0$ and the Hamiltonian collapses to its Hermitian form.  

The symmetry of the Hamiltonian in Eq. \ref{eq:Hamiltonian} captures the $\mathcal{PT}$ symmetry observed experimentally even thought it is not formally $\mathcal{PT}$ symmetric. However, eigenenergies are $\mathcal{PT}$  symmetric, as we demonstrate below. We attributed this $\mathcal{PT}$-symmetry to a chiral exchange interaction\cite{Theiler2025} yielding the $i\alpha\, \sigma\!\cdot\! p$ term, other origins of the non-Hermiticity are possible and can also exhibit $\mathcal{PT}$-symmetry\cite{Volosniev2021,Fransson2025}. Futhermore, non-Hermicity can be generated by introducing external degrees of freedom as mean fields or perturbations\cite{Ding2022},  the following analysis thus treats $\mathcal{PT}$-induced non-Hermiticity as a mechanistically agnostic property, showing first that such $\mathcal{PT}$ Hamiltonians are pseudo-Hermitian with well defined $\eta$ metrics and make use of Gardas thermodynamic equivalence principle\cite{Gardas2016}.

\section{Pseudo-Hermiticity and Dyson Mapping}

For the Hamiltonian in Eq.~\ref{eq:Hamiltonian}, a convenient choice of the similarity transformation, Eq. \ref{Eq. h transform},  is
\begin{equation}
S = \exp\!\left(\tfrac{m\alpha}{\hbar}\, x\!\cdot\!\sigma\right),
\label{Eq.SimilarityTransform}
\end{equation}
with $\tfrac{m\alpha}{\hbar}\in\mathbb{R}$.  The $\mathcal{PT}$-symmetric factor $x\cdot\sigma$ is related to spin displacement\cite{Shi2006} which can be related to physically conserved spin currents. $S$ can thus be understood as the generator of a spin-displacement transformation. The spin-displacement follows from the spin-momentum correlations, and thus chirality makes these correlations non-separable and locked. Crucially, the wavefunctions of Eq.~\ref{eq:Hamiltonian} cannot be factored into a product of spin and spatial components.   

To find $h$ we apply S to  Eq.~\ref{eq:Hamiltonian} and use the Baker--Campbell--Hausdorff formula
\begin{equation}
e^{A} p e^{-A} \;=\; p + [A,p] + \tfrac{1}{2!}[A,[A,p]] + \cdots
\label{Eq:HausdorfCampell}
\end{equation}
where $A=\tfrac{m\alpha}{\hbar}x\cdot\sigma$ and,
\begin{equation}
[A,p] = \tfrac{m\alpha}{\hbar}[x,p]\sigma =  i m \alpha \sigma,
\end{equation}
since $[A,\sigma]=0$, higher commutators in Eq. \ref{Eq:HausdorfCampell} vanish and, 
\begin{align}
S p S^{-1} = p + i m\alpha \sigma.
\label{Eq.p Similarity} \\
S V(x)S^{-1} = V(x).
\end{align}
leading to, 
\begin{align}
h &= \frac{(p + i m \alpha \sigma)^2}{2m} + i\alpha \sigma (p + i m \alpha \sigma)  + V(x)\\
  &= \frac{p^2}{2m} - \frac{d}{2}m\alpha^2 + V(x),
  \label{Eq:h-mappedHamiltonian}
\end{align}

where $d=3$ is the spatial dimension of the Hamiltonian. 

This operator is manifestly Hermitian, $h^\dagger = h$. Hence $H$ is pseudo-Hermitian: it is related by a Dyson map\cite{Dyson1956}, to the Hermitian  Hamiltonian $h$.

The $\eta$-metric operator follows as
\begin{equation}
\eta = S^\dagger S = \exp\left(\tfrac{2m\alpha}{\hbar}x\cdot\sigma\right),
\end{equation}
which is itself Hermitian and positive definite in
\begin{equation}
\mathcal{D}(\eta)={\psi\in L^2 \mid \eta\psi\in L^2}.
\end{equation}
For unbounded domains, square integrability requires $\psi$ to decay superexponential~\cite{Mostafazadeh2013}.
Since $\eta$ is non-periodic in space, the Dyson map cannot be globally defined under periodic boundary conditions, placing the CISS model outside the Bloch-type classification of Ref.~\cite{Kawabata2019} and distinguishes it from the reciprocal-space Kramers-Weyl Fermions in chiral crystals\cite{Chang2018}.
 Under periodic boundary condition, $\eta \to \mathbb{I}$ and the model reduces to its Hermitian limit $\alpha=0$, implying that its topology emerges in real rather than momentum space. This specific Dyson mapping discussed here can thus only be applied to finite systems, consistent with spatial non-reciprocity.

With this metric,  $\eta$, the expectation values of any operator $O$ can be formally computed as
\begin{equation}
\langle O \rangle_\eta = \frac{\langle \psi | \eta O | \psi \rangle}{\langle \psi | \eta | \psi \rangle},
\label{Eq:Normalization}
\end{equation} 
and like with the Hamiltonians, there exist a Dyson-mapping between these inner products or states (Fig. \ref{Fig:Classification}c $\leftrightarrow$ d).

 The pseudo-Hermitian Hamiltonian in Eq.~\ref{eq:Hamiltonian} breaks parity  $\mathcal{P}: x \mapsto -x,  p \mapsto -p,  i \mapsto i,  \sigma \mapsto \sigma $ , time reversal $\mathcal{T}: x \mapsto x,  p \mapsto -p,  i \mapsto -i, \sigma \mapsto -\sigma $,  and $\mathcal{PT}$ because of the structurally chiral potential, $V(x)$. We can define the following four symmetry related Hamiltonians, 

\begin{align}
\mathcal{I}:   &\quad \mathcal{I}H\mathcal{I}^{-1}=\tfrac{p^2}{2m}+i\alpha p\cdot\sigma+V(x) = H_\mathcal{I} , \label{Eq.H_I} \\
\mathcal{P}:& \quad \mathcal{P} H \mathcal{P}^{-1}=\tfrac{p^2}{2m}-i\alpha p\cdot\sigma+V(-x) = H_\mathcal{P}, \\
\mathcal{T}:& \quad \mathcal{T} H \mathcal{T}^{-1}=\tfrac{p^2}{2m}-i\alpha p\cdot\sigma+V(x) = H_\mathcal{T}, \\
\mathcal{PT}: & \quad (\mathcal{PT})H(\mathcal{PT})^{-1}=\tfrac{p^2}{2m}+i\alpha p\cdot\sigma+V(-x). \label{Eq.H_PT}
\end{align}

\begin{figure*}
\centering
\includegraphics[width=0.9\textwidth]{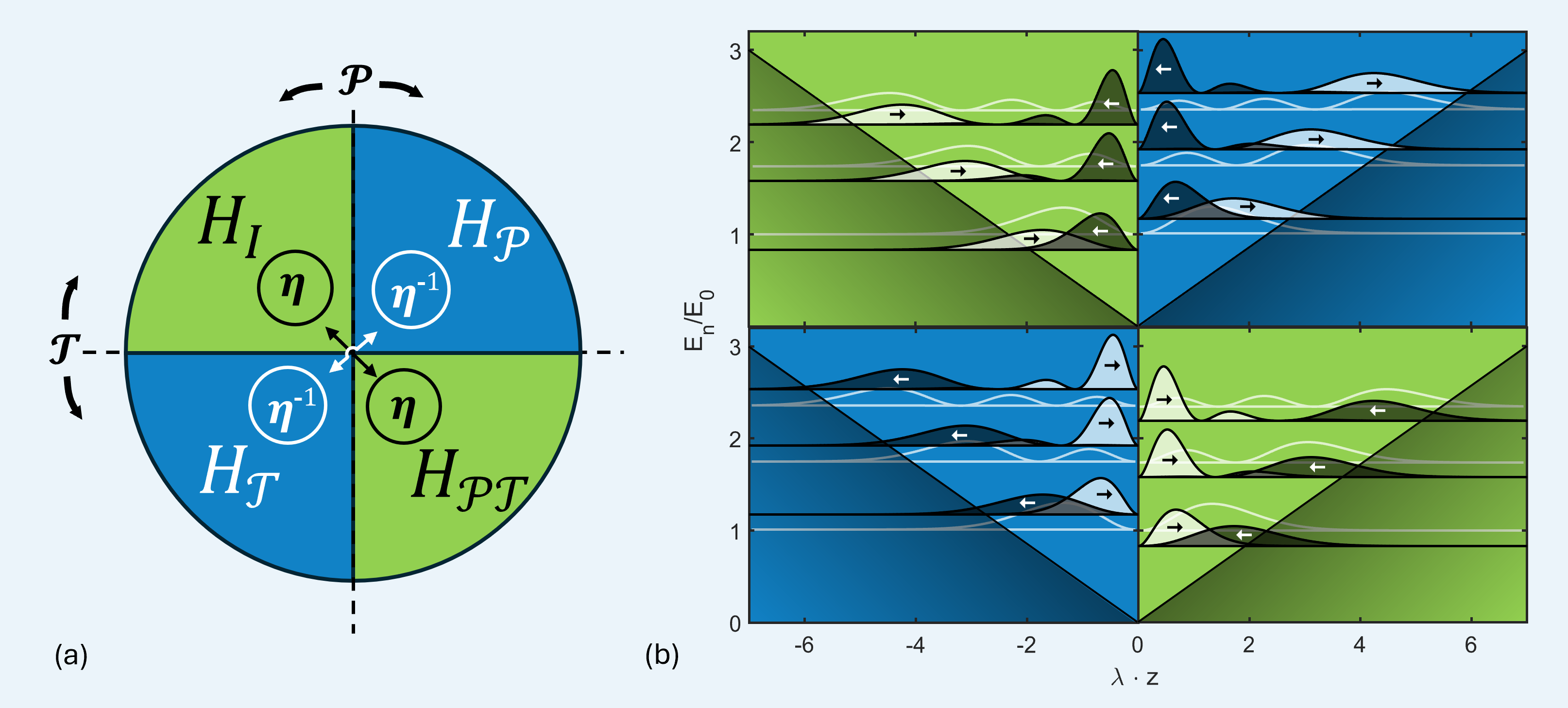}
\caption{Topological separation of the state space in a chiral system into four dynamically disconnected ergodic components: (a) Symmetry relations of the metrics $\eta$ and $\eta^{-1}$ under parity $\mathcal{P}$ and time-reversal $\mathcal{T}$-operation. States with the same metric have degenerate energies but mirror image wavefunctions. (b) Illustration of  Dyson mapping with the same symmetries applied to a triangular potential well problem. The probability distributions is plotted in normalized coordinates $\lambda z$ and staggered according to their normalized energiy $E_n/E_0$. The energy $E_0$ corresponds to the eigenenergy of the ground state of the non-chiral problem $(\alpha=0)$, these eigenstate are plotted as faint white contours. The effect of $\alpha$ is twofold: (1) the eigen energies are shifted by $\alpha^2m/2$ depending upon the chiral sector, (2) spins $\leftarrow$ and $\rightarrow$ are pushed apart to a finite spin-displacement $z\sigma_z$.}
\label{Fig:Sectors}
\end{figure*}

Each of these pseudo-Hermitian Hamiltonians can be mapped to their Hermitian counterparts, through either $\eta$ ,  or $\eta^{-1}$ (Fig. \ref{Fig:Classification}a to \ref{Fig:Classification}b) because,  
\begin{align}
\mathcal{T}\eta \mathcal{T}^{-1} &= \eta^{-1}, \label{Eq:EtaTRS-relation} \\
\mathcal{P}\eta \mathcal{P}^{-1} &= \eta^{-1}, \\
(\mathcal{PT})\eta (\mathcal{PT})^{-1} &= \eta,
\end{align}
Figure~\ref{Fig:Sectors}a show the implications of these symmetry related Hamiltonians: There are four distinct physical sectors each corresponding to one of the Hamiltonians (Eq.\ref{Eq.H_I} to \ref{Eq.H_PT}) related by symmetry and connected with the two metrics $\eta$ and $\eta^{-1}$.  The left ($H_\mathcal{I},H_\mathcal{T}$) and right ($H_\mathcal{P} H_\mathcal{PT}$) pairings form enantiomeric pairs, while the pairs along the diagonal are $\mathcal{PT}$-related pairs (Fig. \ref{Fig:Sectors}) and they share the same metric $\eta$, or $\eta^{-1}$.

Following Callen~\cite{Callen1985}, these four sectors represent the phasespace of permissible states of each system. Within each sector, transitions between microscopic states are allowed, enabling the system to fully explore the available phase space within that sector but cannot cross from one sector to the other. Under these conditions, the ergodic hypothesis applies, and thermal equilibrium corresponds to a uniform sampling of all accessible configurations, while symmetry-forbidden transitions involving $\mathcal{T}$ or $\mathcal{P}$ remain constrained. The states are degenerate for sectors with the same $\eta$ or $\eta^{-1}$ metric, neglecting any parity violations through weak interactions\cite{Quack2022} . 

Additional, we can show that $V(x)$ must be chiral for $\alpha \neq 0$ by contradiction (see Appendix \ref{section: Necessity of a chiral potential for CISS}). Assume a nonchiral potential with mirror symmetry $M$, $V(x)=V(Mx)$, so that $M H_\alpha M^{-1} = H_{-\alpha}$, identifying the green and blue sectors in Fig.~\ref{Fig:Sectors}. However, $H_{\pm\alpha}$ map via distinct similarity transformations to Hermitian counterparts $h_{\pm\alpha} = p^2/2m + V(x) \mp m\alpha^2/2$, which coincide only for $\alpha = 0$. Thus, a finite $\alpha$ requires broken mirror symmetry (Fig.~\ref{Fig:Classification}b, gradient shading), implying that CISS relies solely on the chiral symmetry independent of any specific structural motif~\cite{haque2025} such as previous studied helical-potential models~\cite{Gutierrez2012,Fransson2019,Dalum2019,Dianat2020,Geyer2020,vanRuitenbeek2023,Fransson2025}.

Figure~\ref{Fig:Sectors}b illustrates the solution of Eq. \ref{eq:Hamiltonian} for a triangular-well potential $V(z)=F|z|$ with an infinite wall at $z\le0$ ($z \ge 0$), a configuration which is relevant for CISS experiments involving a ferromagnet or a normal metal interfaced with a chiral material~\cite{Theiler2023,haque2025,Ghosh2020,Fransson2021}. This example demonstrates the application of the Dyson map (see Appendix~\ref{SI:Triangular well}), which transforms the non-Hermitian Hamiltonian $H$ into its Hermitian counterpart $h$. The corresponding eigenvalues and eigenstates read as
\begin{align}
E_{n,\pm \alpha} =& \varepsilon_n \mp \tfrac{1}{2}m\alpha^2 \\
\Psi_{n,s}(z)=&\mathcal{N}_n\, e^{\,s m\alpha z/\hbar}\,\mathrm{Ai}(\lambda z - a_n)\,|\chi_s\rangle,
\end{align}
where $\mathrm{Ai}$ is the Airy function, $\lambda z$ the dimensionless coordinate, and $s=\pm1$ denotes spins parallel or antiparallel to the $z$-axis. The Dyson factor, $e^{sm\alpha z/\hbar}$, exponentially separates the spin components, inducing a spin-dependent spatial displacement, while the entire spectrum is rigidly shifted by $\tfrac{1}{2}m\alpha^2$ corresponding to the different $\eta$-sectors. This uniform energy shift remains unobservable in an isolated chiral material, as level spacings  $\Delta E_{n}$ are preserved for all sectors, but can becomes detectable once the system is coupled to an auxillary system that provides an external energy reference, e.g. (ferromagnetic) interface.

Therefore, it is straightforward to experimentally estimate the size of $\alpha$ wherever an energy scale $|\Delta E_\pm|$ between $\mathcal{T}$- or $\mathcal{P}$-transformed situations is measured,

\begin{equation}
  \alpha = \sqrt{\frac{|\Delta E_\pm|}{m_e}}.
  \label{Eq: alphafromDeltaE}
\end{equation}
these splittings can be from $10^{-3}$ to $10^{0}$ eV\cite{Bloom2024,Aiello2022,Foo2025}, which corresponds to $\alpha= 10^4$ to $10^5$ m/s for the elemental electron's mass $m_{e}$.

The chiral symmetry in the potential $V(\mathbf{x})$ fundamentally distinguishes the present case from altermagnetism~\cite{Smejkal2022}. 
Similar to CISS, altermagnets exhibit a spin texture without relying on conventional spin--orbit coupling\cite{Liu2022}. 
However, altermagnetism requires a well-defined crystal symmetry that locally maps the different magnetic sublattices onto each other, leading to a momentum-dependent spin splitting while preserving overall compensation of the magnetic moments\cite{Smejkal2022}. In contrast, CISS emerges in systems lacking any such spatial symmetry, where the pseudo-Hermitian Dyson map is inherently nonlocal and entangles the spin and spatial coordinates. This nonlocal correspondence replaces the discrete sublattice mapping of altermagnets and links spin polarization directly to structural chirality, suggesting a new magnetic quantum phase: cismagnetism (chirality-induced spin magnetism). The order parameter is $\alpha$ and the transition between the state $\eta$ and $\eta^{-1}$ is smooth, so the transition is second order.

The entanglement between spin and space through the $\eta$-metric places the present model outside the scope of Barron's classification of ``true'' versus ``false'' chirality\cite{LDBarron1981,Barron1987,Barron2020}, which presupposes a conventional Hermitian inner product and the standard strictly local geometric actions of parity and time reversal. In Barron’s framework, two structures are ``truly'' chiral if they cannot be interconverted by time reversal combined with any proper rotation\cite{LaurenceDBarron1986}. 

\section{Detailed Balance and Thermodynamics} 
To find detailed balance conditions we use Wigner's theorem~\cite{Wigner1931}, (depicted in Fig.~\ref{Fig:Classification}c) which states that physical symmetries preserve transition probabilities within each $\eta$-sector. The green $\eta$-sector of Fig.~\ref{Fig:Sectors}a is highlighted in green in Fig. \ref{Fig:Classification}c , an equivalent one exist also for the blue $\eta^{-1}$ sector. Let $|s\rangle$ and $|f\rangle$ denote initial and final states. A unitary operator $U$ satisfies
\begin{equation}
\langle Uf | Us \rangle = \langle f '| s' \rangle, 
\end{equation}
while an antiunitary operator $A$ satisfies
\begin{equation}
\langle Af | As \rangle = \langle f | s \rangle^*.
\end{equation}
Thus, transition probabilities remain invariant,  $|s'\rangle = e^{i\theta}|s\rangle$ is the same pure state under any unitary transformation and:
\begin{equation}
|\langle f' | s' \rangle|^2 = |\langle f | s \rangle|^2|.
\end{equation}
Dresselhaus showed that an odd (even) number of antiunitary operations yields an overall antiunitary (unitary) transformation~\cite{Dresselhaus2008}. Further, to satisfy group properties and the rearrangement theorem, the numbers of $U$ and $A$ operators in a given sector must balance~\cite{Dresselhaus2008}, so that for each $U$ there must be a corresponding $A$.  

Applying the combined operation $AU$ gives
\begin{equation}
\langle AU f | AU s \rangle = \langle f' | s' \rangle^* = \langle s' | f' \rangle,
\end{equation}
corresponding to the microscopically reversed transition. This formalizes the notion that equilibrium states are invariant under the symmetry operations of the system.  We now identify the $U$ and $A$ corresponding to the different sectors of our chiral system. 

In a structurally chiral, spinless molecule (C1 symmetry), the only symmetries are the identity $\mathcal{I}$ and the antiunitary time-reversal $\mathcal{T}$, satisfying the Dresselhaus condition for minimal symmetry and ensuring that chirality alone defines the allowed spin-dependent transitions.

The Hamiltonian in Eq.~\ref{eq:Hamiltonian} is pseudo-Hermitian with respect to the $\eta$-metric, defining a unitary-like symmetry for $H$. Symmetries of $H$ can be obtained by mapping operations from the auxiliary Hermitian system $h$ (Fig.~\ref{Fig:Classification}, c to d). Wigner's theorem requires an antiunitary operator within each $\eta$-sector, but it is not the conventional time-reversal operator $\mathcal{T}$ as that moves from one sector to the other. However $\mathcal{T}$ is usually valid for $h$ since it is Hermitian. 

Unitary symmetries of any chiral molecule are determined by its chiral point group. These unitary elements, together with a suitable antiunitary operator such as $\mathcal{T}$, define the complete set of symmetry operations for the spinful system. In the simplest chiral case, C1, the only unitary symmetry is the identity $\mathcal{I}$, so $\mathcal{T}$ alone provides the required antiunitary operation:

\begin{equation}
\mathcal{T} h \mathcal{T}^{-1} = h^\dagger = h
\end{equation}

Mapping this symmetry back to the original representation (Fig.~\ref{Fig:Classification}, c to d) defines the required antiunitary operator
\begin{equation}
\Theta \equiv S^{-1} \mathcal{T} S.
\label{eq:Theta_def}
\end{equation}
Then applying to our non-Hermitian Hamiltonian

\begin{align}
\Theta H \Theta^{-1}
&= S^{-1}\,\mathcal{T}\,S \; H \; S^{-1}\,\mathcal{T}^{-1}\,S \nonumber\\
&= S^{-1}\,\mathcal{T}\, (S H S^{-1}) \, \mathcal{T}^{-1}\, S \nonumber\\
&= S^{-1}\,\mathcal{T}\, h \, \mathcal{T}^{-1}\, S
\;=\; S^{-1}\, h \, S
\;=\; H,
\end{align}
where we used the properties of pseudo-Hermicity and the similarity transformation. Thus, $\Theta$ is an exact antiunitary symmetry of \(H\):

\begin{equation}
 \Theta H\, \Theta^{-1} = H\ .
\end{equation}

Applying the inverse symmetry mapping from the Hermitian eigenstates, one can show (Appendix \ref{Appendix:ProofReversibility})
\begin{equation}
\langle\phi|_\eta \psi\rangle \;=\; \big\langle \Theta \phi \big|_\eta \, \Theta \psi \big\rangle^{*} = \langle\Theta\psi|_\eta\Theta \phi\rangle .
\end{equation}

That means that $\Theta$ is the symmetry operation that reverses the quantum transitions in the realm of the Hamiltonian Eq. \ref{eq:Hamiltonian} and ensures that transition probabilities $W_{i \to f}^{(\eta)} = |\langle f |_\eta U | i \rangle|^2$ 
remain invariant under  $\Theta$, preserving microscopic reversibility. Consequently, detailed balance retains its canonical form mapped from the Hermitian case\cite{SakuraiQM,Schwabl2008} and is,
\begin{equation}
W_{i \to f}^{(\eta)} \rho_i = W_{f \to i}^{(\eta)} \rho_f,
\end{equation}
with $\rho_i$ and $\rho_f$ denoting equilibrium occupation factors evaluated in the $\eta$-metric space~\cite{Gardas2016,Lifshitz1981}.

In the Hermitian representation, thermal equilibrium is described by the canonical density matrix~\cite{Kubo1992}
\begin{equation}
\rho_h = \frac{e^{-\beta h}}{\operatorname{Tr}[e^{-\beta h}]}.
\end{equation}
Mapping back to the physical chiral system yields
\begin{equation}
\rho_\eta = S^{-1}\rho_h S = \frac{e^{-\beta H}}{\operatorname{Tr}[e^{-\beta H}]}.
\end{equation}

At equilibrium, the density matrix is stationary under the Liouvillian~\cite{Schwabl2008,Kubo1992},
\begin{equation}
\frac{d\rho_\eta}{dt} = \mathcal{L}[\rho_\eta],
\end{equation}
which reduces to the Liouville–von Neumann equation $\dot{\rho}_\eta = -\tfrac{i}{\hbar}[H,\rho_\eta]$ for lossless dynamics, while dissipative systems include additional terms but preserve $\mathrm{Tr}[\rho]=1$. Since $\rho_\eta$ is a function of $H$, one has $[H,\rho_\eta]=0$ which shows that the system can exist in thermodynamic equilibrium. 

The von Neumann entropy,
\begin{equation}
S_{\mathrm{vN}}[\rho_\eta] = -\mathrm{Tr}[\rho_\eta\ln\rho_\eta],
\end{equation}
depends only on the eigenvalues of $\rho_\eta$, which are time-independent. Hence,
\begin{equation}
\frac{dS_{\mathrm{vN}}}{dt}=0,
\end{equation}
and entropy remains constant in pseudo-Hermitian systems in equilibrium irrespective if it is an open or closed system. 

\section{Current operator}

We can define the canonical current (group velocity) operator by
\begin{equation}
J_c \;=\; \frac{\partial H}{\partial p} \;=\; \frac{p}{m} + i\alpha\,\sigma,
\end{equation}
where here \(\sigma\) denotes the Pauli matrix taken along the transport axis. 
Using the similarity map $S$ (Eq.~\ref{Eq.SimilarityTransform})
and the Baker--Campbell--Hausdorff result for $p$ (Eq.~\ref{Eq.p Similarity}), together with \(S\,\sigma\,S^{-1}=\sigma\) (since \([S,\sigma]=0\)), we obtain the similarity-transformed current operator in the Dyson-mapped Hermitian picture,
\begin{align}
J_{h,c} :=& S\,J_c\,S^{-1}
= \frac{1}{m}\,S p S^{-1} + i\alpha\,S\sigma S^{-1} \nonumber\\
=& \frac{1}{m}\big(p - i m\alpha\,\sigma\big) + i\alpha\,\sigma
= \frac{p}{m}.
\end{align}

\section{Spin current operator}

The spin-displacement operator $ x \cdot \sigma$ defines the spin current following Shi \emph{et al}~\cite{Shi2006} together with the Hamiltonian Eq.~\ref{eq:Hamiltonian} we find
\begin{equation}
J_s := \frac{d (x \cdot \sigma)}{dt} = \frac{1}{i\hbar} [x \cdot \sigma, H] = \frac{\sigma \cdot p}{m} + i d \alpha,
\label{spin_current}\end{equation}
where expectation values are taken with respect to the pseudo-Hermitian inner product Eq. \ref{Eq:Normalization}.

Applying the similarity transformation $S$ yields the Dyson-mapped Hermitian counterpart,
\begin{equation}
J_{h,s} = S\,J_s\,S^{-1} = \frac{\sigma \cdot p}{m},
\end{equation}
and the physical expectation value is given by \(\langle J_s\rangle_\eta = \langle\phi|J_{h,s}|\phi\rangle\).

Considering the spin eigenstates along $z$, $\langle\uparrow|J_s|\uparrow\rangle$ and $\langle\downarrow|J_s|\downarrow\rangle$, we find that the individual spin components carry opposite imaginary spin currents. However, their sum is purely real, indicating that the imaginary contribution of Eq.\eqref{spin_current}  implies an internal pairing channel linking the opposite-spin partners. In the Dyson mapping this channel is absorbed into the $\eta$ metric, restoring a Hermitian, real-valued spin current.

The structure of the metric entangles spatial and spin degrees of freedom: the conserved quantities are generated by the pseudo-Hermitian pair \((x\cdot \sigma,\ \tilde p=p + i m\alpha\sigma)\). In particular, \(\tilde p\) is $\tilde{p}^\dagger = \eta \tilde{p} \eta^{-1} $ and hence \([H,x\cdot \sigma]_\eta=0\), which links charge motion to spin displacement. Consequently, transport through the chiral region naturally produces correlated spin and charge flow. Single-electron tunneling operators generally do not preserve the \(\eta\)-symmetry, whereas pair-tunneling operators can; hence pairwise transport restores the pseudo-Hermitian invariance and enforces effective transport in even charge quanta. This provides a symmetry-based rationale for the experimentally observed charge-pairing effects in conductor–chiral–conductor junctions~\cite{Briggeman2025} and for enhanced spin–momentum locking in chiral superconductors~\cite{Sato2025}.

\section{Generalized Onsager Relations for CISS transport}
The standard Onsager-Casimir relations in Hermitian systems rely on three assumptions: 
\begin{enumerate}
    \item The unperturbed system is in thermal equilibrium, satisfying detailed balance and microscopic reversibility\cite{Yang2021}.
    \item The existence of an antiunitary operator $\Theta$ (e.g., time-reversal $\mathcal{T}$), under which observables transform as $\Theta O_i \Theta^{-1} = \epsilon_i O_i$, with $\epsilon_i = \pm 1$ for even/odd operators.
    \item Linear response to small perturbations.
\end{enumerate}
We have shown that the first two conditions can be mapped without restrictions to pseudo-Hermitian systems, whereas the third condition limits the range of applicability. Therefore, we can derive the Onsager relations for spin-charge transport. It has been long known in spintronics, that in limited cases the Onsanger relations can be extended to $\mathcal{T}$-breaking cases under specific symmetries\cite{Gabrielli1996,Bonella2014}, this idea was recently generalized by Huang \emph{et al.}\cite{Huang2025}.

Following this conceptual framework, we can formulate Onsager-like reciprocal relations for the pseudo-Hermitian Hamiltonian in Eq.~\ref{eq:Hamiltonian}, where expectation values are defined with respect to the $\eta$-metric as in Eq.~\ref{Eq:Normalization}.  

Let us define generalized charge, $J^{\mu}_{c}$ , and spin, $J^{\mu}_{s}$,currents and their respective forces:
\begin{align}
J_c^\mu &\equiv \frac{d R_c^\mu}{dt}, & J_s^\mu &\equiv \frac{d R_s^\mu}{dt}, \\
F_c^\nu &\text{ conjugate to } R_c^\nu, & F_s^\nu &\text{ conjugate to } R_s^\nu,
\end{align}
where $R_c^\nu = \sum_j r_{j,\nu}$ is the many-body charge displacement and $R_s^\mu = \sum_j r_{j,\mu} s_{j,z}$ is the many-body spin displacement operator\cite{Shi2006}. The electric force $F_c = -\nabla\mu_c +eF$ can be expressed in terms of the gradient of the chemical potential of the charge $\mu_c$ and the electric field $F$ and the spin force $F_s = -\nabla\mu_s $ can be expressed in terms of the gradient of the chemical potential of the spin $\mu_s$.

The linear response of these fluxes to generalized forces can be written in terms of transport coefficients $L_{ij}$ as
\begin{equation}
\begin{pmatrix} J_c^\mu \\[1mm] J_s^\mu \end{pmatrix}
=
\begin{pmatrix}
L_{cc}^{\mu\nu} & L_{cs}^{\mu\nu} \\
L_{sc}^{\mu\nu} & L_{ss}^{\mu\nu}
\end{pmatrix}
\begin{pmatrix} F_c^\nu \\[1mm] F_s^\nu \end{pmatrix},
\label{Eq:LinearResponse}
\end{equation}
with $L_{cc}, L_{ss}$ describing the conventional charge and spin conductivities, and $L_{cs}, L_{sc}$ the cross-couplings. 

To derive the generalized Onsager relation in the $\eta$-metric, we make use of the antiunitary operator $\Theta$ defined in Eq.~\ref{eq:Theta_def} which as we showed above plays the role of microscopic reversibility in the pseudo-Hermitian system.

To derive the Onsager reciprocity in the pseudo-Hermitian framework, we begin with the Kubo formula (valid in the liner response regime) generalized to the $\eta$-metric~\cite{Kubo1992_2,Callen1985}:
\begin{align}
L_{sc}^{\mu\nu}(\omega) &= i \int_0^\infty dt\, e^{i\omega t} 
\langle [J_s^\mu(t), R_c^\nu(0)] \rangle_\eta, \label{Eq: Kubo_sc}\\
L_{cs}^{\nu\mu}(\omega) &= i \int_0^\infty dt\, e^{i\omega t} 
\langle [J_c^\nu(t), R_s^\mu(0)] \rangle_\eta. \label{Eq: Kubo_cs}
\end{align}
Where expectation values are taken with respect to the $\eta$-inner product.

First, we want to link the above transport terms within a single $\eta$-sector with \(\Theta\) as the antiunitary microscopic-reversibility operator defined in Eq.~\ref{eq:Theta_def}. To make the parity operator under \(\Theta\) explicit we fix the operator content appearing in the correlators:
\begin{align}
R_c^\nu(0) &\equiv r_\nu(0) &\text{(charge displacement)},\nonumber\\
J_c^\nu(t) &\equiv v_\nu(t) &\text{(charge current)},\nonumber\\
R_s^\mu(0) &\equiv x_\mu(0)\, s_\mu(0) &\text{(spin displacement)}, \nonumber\\
J_s^\mu(t) &\equiv v_\mu(t)\, s_\mu(t) &\text{(spin current, spin along \(\mu\))}.\nonumber
\end{align}
Here \(v_\mu=p_\mu/m\) is the velocity operator, \(r_\nu\) the position operator and \(s_i\) the spin operator.

The action of \(\Theta\) on the elementary operators is standard:
\begin{align}
\Theta\, v_\mu(t)\,\Theta^{-1} &= -\,v_\mu(-t),\qquad \\
\Theta\, r_\mu(t)\,\Theta^{-1} &= +\,r_\mu(-t),\qquad  \\
\Theta\, s_i(t)\,\Theta^{-1} &= -\,s_i(-t), 
\end{align}
where the minus signs for \(v\) and \(s_i\) reflect the reversal of velocity and spin under time reversal, and the position operator is even.

From these rules we obtain the parity of the composite operators under \(\Theta\):
\begin{align}
\Theta\, J_s^\mu(t)\,\Theta^{-1}
&= \Theta\big(v_\mu(t)s_\mu(t)\big)\Theta^{-1}
= +\,J_s^\mu(-t),\label{eq:Theta_Js}\\
\Theta\, R_c^\nu(0)\,\Theta^{-1}
&= +\,R_c^\nu(0),\label{eq:Theta_Rc}\\
\Theta\, J_c^\nu(t)\,\Theta^{-1}
&= -\,J_c^\nu(-t),\label{eq:Theta_Jc}\\
\Theta\, R_s^\mu(0)\,\Theta^{-1}
&= -\,R_s^\mu(0).\label{eq:Theta_Rs}
\end{align}

Now consider the two-point correlator appearing in the Kubo formula (all expectation values are \(\eta\)-inner-product expectations):
\[
\langle J_s^\mu(t)\,R_c^\nu(0)\rangle_\eta.
\]
Using the antiunitarity of \(\Theta\) (i.e. \(\langle A\rangle_\eta = \langle \Theta A\Theta^{-1}\rangle_\eta^*\)) and Eqs.~\eqref{eq:Theta_Js}--\eqref{eq:Theta_Rc} we find
\begin{align}
\langle J_s^\mu(t)\,R_c^\nu(0)\rangle_\eta
&= \big\langle \Theta\,J_s^\mu(t)\,R_c^\nu(0)\,\Theta^{-1}\big\rangle_\eta^{*} \nonumber\\
&= \big\langle J_s^\mu(-t)\,R_c^\nu(0)\big\rangle_\eta^{*}.
\end{align}
Next, we reorder the operators inside the complex-conjugated correlator and use 
Eqs.~\eqref{eq:Theta_Jc}--\eqref{eq:Theta_Rs} to relate it to the complementary correlator:

\begin{align}
\big\langle J_c^\nu(t)\,R_s^\mu(0)\big\rangle_\eta \nonumber &= \big\langle \Theta\,J_c^\nu(t)\,R_s^\mu(0)\,\Theta^{-1}\big\rangle_\eta^{*}  \\
&= \big\langle J_c^\nu(-t)\,R_s^\mu(0)\big\rangle_\eta^{*} \\
&= -\big\langle J_s^\mu(-t)\,R_c^\nu(0)\big\rangle_\eta^{*}
\end{align}

Combining the two equalities yields the compact relation between the bare correlators
\begin{equation}
\langle J_s^\mu(t)\,R_c^\nu(0)\rangle_\eta
= -\,\langle J_c^\nu(-t)\,R_s^\mu(0)\rangle_\eta.
\label{eq:correlator_sign}
\end{equation}

Finally, substituting Eq.~\eqref{eq:correlator_sign} into the Kubo expressions (applied separately to the two terms in each commutator). One obtains immediately
\begin{equation}
  L_{sc}^{\mu\nu}(\omega) \;=\; -\,L_{cs}^{\nu\mu}(\omega) \;,
\end{equation}
which defines the reciprocity under \(\Theta\) within a given \(\eta\)-sector.

Since $H_\mathcal{I}$ and $H_\mathcal{PT}$ share the same metric $\eta$, the transport coefficients can be mapped exactly, although they describe physically different enantiomeric pairs, this also applies identically to the pair $H_\mathcal{P}$ and $H_\mathcal{T}$ with the metric $\eta^{-1}$. However, since all four sectors are related by symmetry, therefore we can also find a connection between $L_{cs}^\eta$ and $L_{cs}^{-\eta}$ that can be derived via the Kubo formalism and results in the generalized Onsager relation within each enantiomer. 

Since these metrics are related by time reversal (Eq. \ref{Eq:EtaTRS-relation}), expectation values in the two $\eta$-sectors can be connected using the definition of an expectation value (Eq. \ref{Eq:Normalization}). 
For any operator $O$, antiunitarity of $\mathcal{T}$ implies
\begin{equation}
\langle O \rangle_{\eta^{-1}}
= \big\langle \mathcal{T} O \mathcal{T}^{-1} \big\rangle_\eta^{*},
\label{eq:eta_eta_inv_expectation}
\end{equation}
where the complex conjugation arises from the antiunitary nature of $\mathcal{T}$.

Applying Eq.~\eqref{eq:eta_eta_inv_expectation} to the Kubo expression in Eq.~\ref{Eq: Kubo_sc}, we find
\begin{align}
\langle [J_c^\nu(t), R_s^\mu(0)] \rangle_{\eta^{-1}}
&= 
\Big\langle \mathcal{T}\,[J_c^\nu(t),R_s^\mu(0)]\,\mathcal{T}^{-1} \Big\rangle_\eta^{*}.
\end{align}
Time reversal changes the sign of both the charge current and the spin displacement operator,
\begin{equation}
\mathcal{T}\,J_c\,\mathcal{T}^{-1} = -J_c, \qquad 
\mathcal{T}\,R_s\,\mathcal{T}^{-1} = -R_s,
\end{equation}
so that the commutator is invariant under $\mathcal{T}$:
\begin{equation}
\mathcal{T}\,[J_c^\nu(t),R_s^\mu(0)]\,\mathcal{T}^{-1}
= [J_c^\nu(t),R_s^\mu(0)].
\end{equation}
Inserting this into the Kubo formula for $L_{cs}$ yields
\begin{align}
L_{cs}^{\nu\mu}(\omega;\eta^{-1})
&= i\!\int_0^{\infty}\!dt\, e^{i\omega t}\,
\langle [J_c^\nu(t),R_s^\mu(0)]\rangle_{\eta^{-1}} \nonumber\\
&= i\!\int_0^{\infty}\!dt\, e^{i\omega t}\,
\Big(\langle [J_c^\nu(t),R_s^\mu(0)]\rangle_\eta\Big)^{*} \nonumber\\
&= -\Big(i\!\int_0^{\infty}\!dt\, e^{-i\omega t}\,
\langle [J_c^\nu(t),R_s^\mu(0)]\rangle_\eta \Big)^{*}.
\label{Eq.Eta-Relation_Kubo}
\end{align}
Recognizing the Kubo integral at frequency $-\omega$, we obtain the exact frequency-dependent relation by comparing Eq.\ref{Eq.Eta-Relation_Kubo} and \ref{Eq: Kubo_sc}
\begin{equation}
L_{cs}^{\nu\mu}(\omega;\eta^{-1}) 
= -\big(L_{cs}^{\nu\mu}(-\omega;\eta)\big)^{*}.
\label{Eq:Lcs_general_relation}
\end{equation}
In the static limit $\omega\to 0$, where $L_{cs}$ is real, Eq.~\eqref{Eq:Lcs_general_relation} reduces to
\begin{equation}
L_{cs}^{\nu\mu}(\eta^{-1}) = -L_{cs}^{\nu\mu}(\eta),
\end{equation}
confirming that the cross-coupling coefficients of $\eta$- and $\eta^{-1}$-sectors differ by a sign.

While the off-diagonal coefficients $L_{cs}$ and $L_{sc}$ change sign between sectors related by $\eta \leftrightarrow \eta^{-1}$, the diagonal coefficients $L_{cc}$ and $L_{ss}$ remain invariant for all four non-Hermitian Hamiltonians $H_\mathcal{I}, H_\mathcal{P}, H_\mathcal{T}, H_\mathcal{PT}$. This invariance holds regardless of the parity of $V(x)$, reflecting the fact that charge-charge and spin-spin responses do not depend on the choice of metric $\eta$.

\begin{table}[h!]
\centering
\caption{Frequency-dependent transport coefficients $L(\omega)$ for the four pseudo-Hermitian sectors. Diagonals show the same transport coefficients because they are sharing the same $\eta$-metric.}
\begin{tabular}{c|c|c}
 & $\mathcal{I}$ & $\mathcal{P}$ \\ \hline
$\mathcal{I} $ & 
$\begin{pmatrix} L_{cc} & L_{cs}(\omega) \\[1mm] -L_{cs}(\omega) & L_{ss} \end{pmatrix}$ & 
$\begin{pmatrix} L_{cc} & -L_{cs}(-\omega)^* \\[1mm] L_{cs}(-\omega)^* & L_{ss} \end{pmatrix}$ \\ \hline
$\mathcal{T} $ & 
$\begin{pmatrix} L_{cc} & -L_{cs}(-\omega)^* \\[1mm] L_{cs}(-\omega)^* & L_{ss} \end{pmatrix}$ & 
$\begin{pmatrix} L_{cc} & L_{cs}(\omega) \\[1mm] -L_{cs}(\omega) & L_{ss} \end{pmatrix}$ \\
\end{tabular}
\label{tab:transport_eta_simplified}
\end{table}

The generalized Onsager relations derived above have several important consequences for spin-charge coupled transport in pseudo-Hermitian systems. 

First, the antiunitary $\Theta$-symmetry ensures that the off-diagonal transport coefficients, $L_{cs}$ and $L_{sc}$, are antisymmetric. This antisymmetry implies that spin-to-charge or charge-to-spin conversion occurs without generating additional entropy beyond that associated with the diagonal conductivities $L_{cc}$ and $L_{ss}$. This can be easily proven by considering that $\dot S =\frac{1}{T} \sum_i J_i F_i =\frac{1}{T}  X^T L X$. In other words, the cross-coupling represents a dissipationless transduction channel for spin and charge.  This is a direct consequence of the entangled spin and charge degrees of freedom capture in the non-Hermitian Hamiltonian and embedded in the $\eta$ metric. If the spin is driven out of equilibrium charge must componensate and if the charge is driven out of equilibrium then the spin must compensate.  

Second, the mapping between $\eta$ and $\eta^{-1}$ metrics, expressed in Eq.~\eqref{Eq:Lcs_general_relation}, reveals a fundamental constraint on the frequency-dependent response: the cross-coupling in one metric is related to the complex conjugate and negative frequency of the other. In the static limit $\omega \to 0$, this reduces to a simple sign flip, highlighting a universal relation between the four pseudo-Hermitian sectors $H_\mathcal{I}, H_\mathcal{P}, H_\mathcal{T}, H_\mathcal{PT}$. In contrast, the diagonal coefficients $L_{cc}$ and $L_{ss}$ remain invariant across all sectors, reflecting the fact that pure charge and spin responses are independent of the choice of the $\eta$-metric.

Third, $\mathcal{PT}$-symmetry constrains the functional form of the induced chemical potentials under simultaneous reversal of magnetization and enantiomer. This symmetry explains why experimental observations of inverse CISS (ICISS) show even behavior under combined flips (i.e., changing both magnetic polarization and stereochemical configuration), while individual flips (i.e., only changing magnetic polarization or stereochemical configuration) do not correspond to any symmetry of the system. Consequently, $\mathcal{PT}$ provides a fundamental symmetry principle that can be used to design and interpret CISS experiments that probe the spin-to-charge interconversion.

Fourth, in the pseudo-Hermitian Hamiltonian of Eq.~\eqref{eq:Hamiltonian}, the spin couples exclusively along the momentum direction via the term $i\alpha\, \sigma\!\cdot\! p$. This enforces that the spin response is strictly collinear with the transport direction. Within the $\eta$-metric, expectation values of spin currents inherit this alignment, $\langle J_s^\mu \rangle_\eta \propto \langle \sigma^\mu \cdot p^\mu \rangle_\eta$, so that off-diagonal spin-charge coefficients vanish, $L_{cs}^{\mu\nu} = 0$ for $\mu \neq \nu$, while the longitudinal component remains finite, $L_{cs}^{\mu\mu} \neq 0$.

Finally, the pseudo-Hermitian framework naturally integrates frequency dependence, allowing one to predict both static and dynamical transport properties. The combination of dissipationless cross-couplings, metric-dependent sign relations, and symmetry-enforced constraints provides a unified description of spin-charge transport in chiral conductors. These features make pseudo-Hermitian Onsager reciprocity a powerful tool for understanding and exploiting CISS phenomena in both equilibrium\cite{Theiler2023,BenDor2017} and driven regimes\cite{dong2025}.

\section{Scattering and Bardarson’s Theorem}

Bardarson’s theorem~\cite{Bardarson2008} proofs that in $\mathcal{T}$-symmetric two-terminal systems with $\mathcal{T}^2=-1$, the unitary scattering matrix is antisymmetric and transmission eigenvalues are Kramers-degenerate, precluding net spin-polarized currents. However, this argument relies on (i) Hermiticity and $\mathcal{T}$-invariance of the Hamiltonian, (ii) the existence of Kramers pairs $|n\rangle$ and $i\sigma_y \mathcal{T}|n\rangle$ , and (iii) implicit spatial unitary symmetries that map the system onto itself\cite{Matityahu2016}.

In the pseudo-Hermitian framework, the role of $\mathcal{T}$ is taken over by the antiunitary operator $\Theta$, with expectation values defined via the $\eta$-metric. Applying Bardarson's argument to the underlying Hamiltonian $h$ with $\Theta$ symmetry, we find that the diagonal transport coefficients $L_{cc}$ and $L_{ss}$ are constrained to be identical across sectors connected by $\Theta$ ($\eta \leftrightarrow \eta^{-1}$), because $\Theta$ enforces the degeneracy of the underlying transmission eigenvalues. Consequently, the diagonal elements of the transport matrix cannot differ between the four pseudo-Hermitian sectors $H_\mathcal{I}, H_\mathcal{P}, H_\mathcal{T}, H_\mathcal{PT}$.  

In contrast, the off-diagonal coefficients $L_{cs}$ and $L_{sc}$ are metric-dependent and transform under $\Theta$ according to the generalized pseudo-Hermitian Onsager relations, allowing them to change sign or acquire frequency dependence between sectors. Thus, while $\Theta$-generalized Bardarson constraints preserve the invariance of charge-charge and spin-spin responses, they do not forbid spin-polarized transport mediated by chiral cross-couplings. This could help to settle the debate of linear transport in CISS\cite{Dalum2019,Evers2022,Wolf2022,Yang2021,Utsumi2020,Foo2025,Aharony2025}, on how to interpret Bardarson's theorem to CISS transport experiments.

\section{Experimental observables}
The transport coefficients of equation Eq. \ref{Eq:LinearResponse} can be used, to compare on a quantitative level different observables with each other: 

A CISS magnetoresistance experiment probes the ratio of spin-resolved conductances\cite{haque2025}. In the linear-response regime, charge and spin currents are related to their conjugate forces via Eq.~\ref{Eq:LinearResponse}, where \(J_c = J_\uparrow + J_\downarrow\) denotes the charge flux and \(J_s = J_\uparrow - J_\downarrow\) the spin flux along a fixed quantization axis along the charge transport axis.  
The generalized forces \(F_c\) and \(F_s\) correspond to charge and spin driving fields, respectively; under a standard two-terminal voltage bias, one has \(F_c \propto V\) and \(F_s = 0\) in the absence of a spin bias.  

When only a voltage is applied (\(F_s=0\)), the experimentally measured spin polarization is defined as
\begin{equation}
P \equiv \frac{J_\uparrow - J_\downarrow}{J_\uparrow + J_\downarrow}
      = \frac{J_s}{J_c}
      = \frac{L_{sc}}{L_{cc}}.
\end{equation}

Thus, the observed spin polarization reflects the ratio of the spin--charge coupling \(L_{sc}\) to the charge conductivity \(L_{cc}\), consistent with the constant experimental ratio. For the opposite enantiomer, the polarization reverses sign (\(P \rightarrow -P\)), as shown in Table~\ref{tab:transport_eta_simplified}. This relation, however, holds strictly only in the absence of interfacial spin forces or large perturbations.

In the spinterface model\cite{Sarkar2025}, both a spin polarization and a bare $I(V)$ curve are required as input parameters. Although the rational behind the mechanism is different, our dissipative-free framework is equivalent to the spinterface-model\cite{Alwan2021,Alwan2023,Alwan2024}, there a $I(V)$ of a non-magnetic interface and a spin polarization is need as input : $I(V)\!\propto\!L_{cc}$ is the nonmagnetic transport response, while spin dependence enters through a correction proportional to $L_{sc}$, explaining quantitative good fitting agreements. 

Alternatively, the correlated open-system approaches developed by Fransson account for electron–electron interactions by coupling an interacting bath to electron states, thereby reproducing many-body effects~\cite{Fransson2019,Fransson2020a,Fransson2021,Fransson2025}. From a mechanistic standpoint, this concept bears close resemblance to the present framework and may, under specific conditions, be regarded as formally equivalent~\cite{Michishita2020}, wherein electronic correlations are effectively represented by a non-Hermitian contribution to the Hamiltonian. A principal limitation of the non-equilibrium Green’s function formulation, however, lies in its intrinsic dependence on bath coupling, which obscures the identification of equilibrium properties such as cismagnetism.

At a chiral–ferromagnet interface, the entanglement between charge and spin currents implies that a spin voltage can induce a charge response. Within linear response (Eq.\ref{Eq:LinearResponse}) and under open-circuit conditions ($J_c = 0$, $J_s = 0$ ), the induced voltage needs to compensates the spin-driven current $F_c = - L_{cc}^{-1} L_{cs} F_s$, leading to a measurable voltage difference as predicted.

\begin{equation}
    \Delta V = \frac{\Delta \mu_c}{e} =  \frac{L_{sc}}{L_{cc}} \frac{\Delta \mu_s}{e}
\end{equation}
Upon reversing the magnetization of the ferromagnet, the spin voltage $\Delta \mu_s$ changes sign, and hence the induced charge voltage also reverses. Altough already theoretically modelled\cite{Fransson2021} or described as a change in tunnel-barrier height\cite{Adhikari2023,haque2025}, our framework provides a closed expression for the observed magnetization-dependent voltage at chiral–ferromagnet interfaces\cite{Abendroth2019, Ghosh2020,Theiler2023}.

\section{Discussion}

The framework presented here provides a coherent resolution of several long-standing conceptual tensions in the study of chirality-induced spin selectivity (CISS)\cite{Aiello2022,Evers2022,Foo2025}. The framework presented here describes a picture of CISS where only electron correlations and chirality are needed inorder to generate CISS observables, supporting previous proposals in this direction\cite{Naaman2020,Chiesa2025,Fransson2021,Fransson2019}. It reconciles the apparent contradictions between thermodynamic reciprocity, microscopic reversibility, and the observation of equilibrium spin polarization and time-reversal breaking in experiments. By extending quantum mechanics into a pseudo-Hermitian, $\mathcal{PT}$-symmetric formulation, the theory establishes a thermodynamically consistent foundation for describing correlated, structurally chiral systems with spin.

\textit{First}, the present framework clarifies the relationship between microscopic reversibility, time-reversal symmetry, and the emergence of spin polarization in equilibrium. By distinguishing the standard time-reversal symmetry $\mathcal{T}$ from the generalized antiunitary symmetry $\Theta$ of the Hamiltonian, we demonstrate that equilibrium spin polarization is fully compatible with detailed balance and the second law of thermodynamics using Wigner's theorem.  
The pseudo-Hermitian formulation elevates the role of chirality: rather than being a geometric decoration without physical consequences, chirality becomes encoded in the non-local structure of the physical inner product, defining distinct $\eta$ and $\eta^{-1}$ sectors that are thermodynamically different.  
In this sense, the observed spin polarization is a manifestation of non-local correlations enforced by $\eta$, measurable as exceptionally strong electron pairing\cite{Briggeman2025}.  
This view naturally explains the existence of static spin textures that are monopole-like~\cite{Zhu2024} and consistent with recent experimental reports of spin orientation collinear with charge transport~\cite{Nakajima2023,Moharana2025,dong2025}.

 The pseudo-Hermitian framework reveals that electronic correlations emerge as a consequence of structural chirality. The metric operator $\eta$ couples spatial and spin degrees of freedom, giving rise to a chirality-driven spin–displacement order that we term the cismagnetic phase.  Unlike a conventional magnetic order that minimizes an exchange energy, the cismagnet emerges from the topological structure of correlations encoded in the $\eta$-metric. Because the order is only defined in finite systems in real space rather than reciprocal space, periodic boundary conditions may fail to capture CISS and defy conventional band structure analysis\cite{dong2025}, as discussed in Ref.~\cite{Kunst2018}.  In this sense, CISS is neither a pure interface nor a bulk effect: it represents the experimental manifestation of the cismagnetic ordering.

The analysis also highlights a practical route to extend electronic-structure simulations beyond conventional Hermitian frameworks. Hermitian single-particle models lack built-in correlations and thus cannot account for the experimentally observed non-reciprocal spin transport. The Dyson mapping introduced here overcomes this limitation by mapping the pseudo-Hermitian Hamiltonian to a Hermitian one with a constant energy shift representing the correlation energy. This mapping allows direct augmentation of \emph{ab initio} methods such as density functional theory, enabling calculation of wavefunctions consistent with the present analytical results\cite{Burton2019}. 

Importantly, the inverse Dyson mapping is defined by a single parameter $\alpha$ that is fixed by an agnostic correlation mechanism that results in a spin displacement $\sigma\cdot x$. Since the metric $\eta$ commutes with the potential, no specific potential need to be assumed for the model, except that any mirror symmetry is absent addressing the role of specific geometries\cite{Foo2025}. Therefore, we can go beyond assuming a helical potential model~\cite{Gutierrez2012,Fransson2019,Dalum2019,Dianat2020,Geyer2020,vanRuitenbeek2023,Fransson2025}.  Additionally, the presented framework does not need to assume spin-orbit coupling. For conventional spin-orbit coupling, the symmetry of the spin-orbit interaction does not match the required transformations under $\mathcal{P}$ and $\mathcal{T}$ observed in experiments, requiring an additional $\mathcal{T}$-breaking mechanism such as dissipation\cite{Aharony2025,Volosniev2021} in order to reconcile with experiments. Such approaches differ from the approach presented here that explains CISS as an equilibrium phenomena. Thus, the framework presented here could provide a route to resolve the mismatch of the apparent spin-orbit coupling strength needed to explain CISS experiments with the established conventional spin-orbit coupling values in light elements\cite{Foo2025}. Instead, the chiral–charge interaction strength can be described by a scalar encoded in $\alpha$ that explicitly depends on the electric toroidal monopole $G_0$~\cite{Kishine2022,Inda2024,Kusunose2024}, avoiding the need to introduce an ad hoc ``chiral axis''~\cite{Sarkar2025}. More research is needed to elucidate the microscopic origin of the spin–displacement correlation $x \cdot \sigma$ and to establish the functional relationship between $\alpha$ and $G_0$.  Such studies will be essential to formulate design principles for maximizing the chiral–spin interaction and to connect first-principles calculations with the framework developed here. 

\textit{Second}, extending linear response theory to pseudo-Hermitian systems reveals that the Onsager–Casimir relations can be generalized to chiral–spin systems when expectation values are redefined with the $\eta$-inner product. Conventional Onsager-Casimir relations are indeed not valid for CISS, but they can be generalized, proving their generality\cite{Yang2021,Tirion2024}. This leads to antisymmetric charge–spin response coefficients that naturally explain the dissipationless character of spin–charge transduction in chiral conductors. In this framework, we can answer the question is CISS a spin-filter vs polarizer?\cite{Aiello2022,Foo2025}: CISS emerges not as a spin-filtering\cite{Kishine2022,Utsumi2020} phenomenon where one spin channel is dissipatively suppressed. Rather, the framework has a spin-polarizing transduction process\cite{Wolf2022,Vensaus2025} in which the charge and spin currents remain phase-correlated and conserve energy. However, it can be perceived as a spin filter in the case where dissipation is active through $L_{ss}$ or $L_{cc}$. Notably, the inverse spin-to-charge process (inverse CISS) has been recently reported~\cite{Sun2024,dong2025}. From an experimental standpoint, this implies that finite spin and charge currents can coexist with zero entropy production, consistent with reports of equilibrium spin polarization and open-circuit spin voltages in molecular devices. Moreover, the formalism offers a quantitative method to determine effective coupling coefficients directly from measurable quantities, providing a unified framework for comparing distinct families of CISS experiments. The spin displacement should also be measurable as a quantum inductance as a result in a change of magnetic permeability in response to an electric polarization change. 

\textit{Third}, our analysis extends Bardarson’s theorem\cite{Bardarson2008} to chiral systems. In contrast to conventional Hermitian, $\mathcal{T}$-symmetric conductors, parity is explicitly broken in chiral structures, and spin and spatial degrees of freedom are entangled. As a result, the pairing of transmission eigenvalues assumed in the original theorem fails and applies just for $\eta$-metric pairs, and spin-polarized transport becomes possible even in two-terminal geometries.  This reframing resolves a long-standing conceptual deadlock, invalidating therefore this no-go argument\cite{Utsumi2020,Utsumi2022,Aharony2025}, reconciling the experimental observation of (equilibrium) spin polarization in two-terminal devices\cite{Hautzinger2024,Safari2024}.

Temperature dependence\cite{Sarkar2025} has not been explicitly addressed here, but the pairing mechanism suggests that CISS might exhibit remarkable resistance to thermal dephasing. The linearized transport equations underline the intricate temperature dependencies already noted by others. For instance, in magnetoresistance devices, the observed ratio depends not only on the temperature dependence of spin–charge conversion but also on charge conduction itself. Thus, the present framework may help disentangle overlapping temperature effects that do not originate from chirality.  

The pairing of electrons is also of potential relevance to biological systems~\cite{Aiello2022,Naaman2021}, as it provides a mechanism for stable correlations at room temperature that may support quantum biological processes such as information transfer~\cite{Goren2025}, energy-transfer entanglement~\cite{Buckel2018}, and chiral recognition~\cite{Kumar2017}.  Such processes could have been essential in the emergence of biological homochirality~\cite{Ozturk2022}.  More broadly, the non-local nature of the metric operator implies that effective spin correlations are spatially extended, hinting at a mechanism for spin coherence in chiral materials and biomolecules\cite{Naaman2021}.  This motivates future experiments to probe the spatial structure of spin polarization and its dependence on molecular geometry and device architecture. 

Finally, the characteristic timescales associated with chiral spin–charge correlations are of particular interest for quantum applications operating at elevated temperatures. Should the present framework confirm that electron–electron correlations are responsible for CISS, this would imply extremely fast transduction speeds~\cite{dong2025}. Recent ultrafast studies have indeed reported chiral signatures on attosecond timescales~\cite{Han2025}. Using Heisenberg’s uncertainty principle, the characteristic time can be estimated as $\tau \propto \hbar / (\alpha^2 m) \approx 7~\mathrm{fs}$, corresponding to a characteristic frequency of about $150~\mathrm{THz}$. Therefore, CISS effects should be detectable down to the femtosecond range. 

In agreement with Refs.~\cite{Sarkar2025,Bloom2024}, more quantitative comparisons between theory and experiment are required.  
We anticipate that the introduction of generalized Onsager-like relations and the analytical formalism developed here will enable explicit modeling in future work.  
Further modeling efforts are underway to relate the order parameter $\alpha$ directly to measurable Onsager coefficients,  enabling comparison between static, transport, and photo-excitement experiments beyond Eq.~\ref{Eq: alphafromDeltaE}.  

Taken together, these insights position pseudo-Hermitian quantum mechanics as a unifying lens through which the experimental richness of CISS can be interpreted as a thermodynamic stable cismagnetic quantum phase.

\begin{acknowledgments}
The authors acknowledge useful discussions during a CISS Theory Workshop held at NREL in May 2025, in particular with Mark van Schilfgaarde, Jonas Fransson, Yonathan Dubi, Martin van Horn and Eva Theiler for their insightful comments and fruitful discussions. 

\textbf{Funding: }This project was supported as part of the Center for Hybrid Organic Inorganic Semiconductors for Energy (CHOISE) an Energy Frontier Research Center funded by the Office of Basic Energy Sciences, Office of Science within the U.S. Department of Energy. This work was authored in part by NREL under Contract No. DE-AC36-08GO28308 to DOE. The views expressed in the article do not necessarily represent the views of the DOE or the U.S. Government.\\
\textbf{Competing interests:} The authors declare no competing interests. \\
\textbf{Data availability:} The main data that support the findings of this study are available in this article. Additional data are available from the corresponding authors upon reasonable request. \\

\end{acknowledgments}

\appendix

\section{Necessity of a chiral potential for CISS}
\label{section: Necessity of a chiral potential for CISS}
Assume the potential is non-chiral, i.e.\ there exists a spatial reflection/rotation \(R\) (for brevity denoted by the operator \(\mathcal{P}_R\)) such that
\begin{equation}
V(Rx)=V(x) \qquad\Longrightarrow\qquad \mathcal{P}_R V(x)\mathcal{P}_R^{-1}=V(x).
\label{eq:V_symmetry}
\end{equation}
Consider the family of non-Hermitian Hamiltonians (Eq.~\eqref{eq:Hamiltonian}) parameterized by \(\alpha\):
\begin{equation}
H_\alpha = \frac{p^2}{2m} + i\alpha\,\sigma\!\cdot\! p + V(x).
\end{equation}
Under the spatial operation \(\mathcal{P}_R\) we have \(p\mapsto -p\) (for parity-like \(R\)) and, using \eqref{eq:V_symmetry},
\begin{equation}
\mathcal{P}_R H_\alpha \mathcal{P}_R^{-1} = \frac{p^2}{2m} - i\alpha\,\sigma\!\cdot\! p + V(x) \equiv H_{-\alpha}.
\label{eq:P_maps_H}
\end{equation}

For each \(\alpha\) define the similarity transform (Eq.~\eqref{Eq.SimilarityTransform})
\begin{equation}
S_\alpha = \exp\!\Big(-\frac{m\alpha}{\hbar}\,x\!\cdot\!\sigma\Big),
\end{equation}
which yields the Hermitian image
\begin{equation}
h_\alpha \equiv S_\alpha H_\alpha S_\alpha^{-1} = \frac{p^2}{2m} + \frac{d}{2}m\alpha^2 + V(x).
\end{equation}
Because \(V(x)=V(Rx)\) the two Hermitian images coincide:
\begin{equation}
h_\alpha = h_{-\alpha}.
\label{eq:h_equality}
\end{equation}
The corresponding metrics are
\begin{equation}
\eta_\alpha = S_\alpha^\dagger S_\alpha,\qquad \eta_{-\alpha}=S_{-\alpha}^\dagger S_{-\alpha}.
\end{equation}
Noting \(S_{-\alpha}=S_\alpha^{-1}\) we obtain
\begin{equation}
\eta_{-\alpha} = \big(S_\alpha^{-1}\big)^\dagger S_\alpha^{-1} = \eta_\alpha^{-1}.
\label{eq:eta_inverse}
\end{equation}

Now assume \(\alpha\neq0\). Then \(S_\alpha\) is non-unitary and therefore \(\eta_\alpha\neq\mathbb{I}\) and \(\eta_{-\alpha}=\eta_\alpha^{-1}\neq\eta_\alpha\).
However, by \eqref{eq:P_maps_H} and \eqref{eq:h_equality} the two non-Hermitian Hamiltonians \(H_\alpha\) and \(H_{-\alpha}\) are related by the exact spatial symmetry \(\mathcal{P}_R\) and share the same Hermitian image \(h\). Physical observables computed in the two sectors must therefore be identical under the symmetry \(\mathcal{P}_R\). This identification requires the same physical inner product (metric) in both sectors, i.e. \(\eta_{-\alpha}=\eta_\alpha\). Combining this with \eqref{eq:eta_inverse} gives
\begin{equation}
\eta_\alpha = \eta_\alpha^{-1} \quad\Longrightarrow\quad \eta_\alpha^2=\mathbb{I}.
\end{equation}
Since \(\eta_\alpha\) is positive definite, the last equality implies \(\eta_\alpha=\mathbb{I}\). Hence \(S_\alpha^\dagger S_\alpha=\mathbb{I}\), so \(S_\alpha\) is unitary. But
\begin{equation}
S_\alpha = \exp\!\Big(-\frac{m\alpha}{\hbar}\,x\!\cdot\!\sigma\Big)
\end{equation}
is unitary only if \(\frac{m\alpha}{\hbar}\) is purely imaginary; for real \(\alpha\) this forces \(\alpha=0\), contradicting the assumption \(\alpha\neq0\).

Therefore the assumption that a non-chiral potential \(V(x)=V(Rx)\) can coexist with \(\alpha\neq0\) leads to a contradiction. We conclude that nonzero chiral coupling \(\alpha\) requires a structurally chiral potential (no mirror or parity symmetry). Equivalently, if \(V(x)\) admits a parity/mirror symmetry then \(\alpha\) must vanish.

\section{Triangular well}
\label{SI:Triangular well}
Without loss of generality, the Hamiltonian in Eq.~\ref{eq:Hamiltonian}  can be reduced to a 1D problem with momentum along \(z\) (so \(\sigma\!\cdot\!p\to\sigma_z p_z\)) and a linear potential
\begin{equation}
V(z)=F\,z,\qquad z>0,
\end{equation}
with an infinite wall at \(z\le0\), apply the Dyson map of Eq.~\ref{Eq.SimilarityTransform}.
As shown in the main text (Eq.~\ref{Eq:h-mappedHamiltonian}), the similarity transform maps \(H\) to the Hermitian operator \(h\)
Thus the spectral problem for \(H\) is equivalent to that of \(h\). Denote by \(\mathcal{E}\) an eigenvalue of \(h\) and by \(\varphi(z)\) the corresponding spin-scalar eigenfunction; split the energy as
\begin{equation}\label{eq:energy_split}
\mathcal{E} = \varepsilon \mp \frac{1}{2}m\alpha^2,
\end{equation}
where the sign of $\alpha$ defines the different enantiomers of the 3D-chiral potential and \(\varepsilon\) satisfies the pure triangular-well (Hermitian) problem
\begin{align}\label{eq:hermitian_triangular}
\left[-\frac{\hbar^2}{2m}\frac{d^2}{dz^2} + F z\right]\varphi(z) = \varepsilon\,\varphi(z)\\
\qquad z>0,\quad \varphi(0)=0,\ \varphi(z)\xrightarrow{z\to\infty}0.\nonumber
\end{align}

Introduce the scale
\begin{equation}
\lambda \equiv \Big(\frac{2mF}{\hbar^2}\Big)^{1/3},
\end{equation}
and the dimensionless coordinate \(\xi=\lambda z - a\). The square-integrable solution of Eq.~\eqref{eq:hermitian_triangular} is the Airy function,
\begin{equation}\label{eq:phi_Ai}
\varphi_n(z)=\mathcal{N}_n\,\mathrm{Ai}(\lambda z - a_n),\qquad n=1,2,\dots,
\end{equation}
with the quantization condition
\begin{equation}\label{eq:airy_zero_condition}
\mathrm{Ai}(-a_n)=0,\qquad a_1\approx2.3381,\ a_2\approx4.0879,\ldots,
\end{equation}
and energies
\begin{equation}\label{eq:epsilon_n}
\varepsilon_n = \Big(\frac{\hbar^2}{2m}\Big)^{1/3} F^{2/3}\, a_n.
\end{equation}
Combining with Eq.~\eqref{eq:energy_split} yields the full eigenvalues of \(H\),
\begin{equation}\label{eq:E_n_final}
E_n = \frac{1}{2}m\alpha^2 + \Big(\frac{\hbar^2}{2m}\Big)^{1/3} F^{2/3}\, a_n\,,\qquad n=1,2,\dots\
\end{equation}

Eigenfunctions of the original (non-Hermitian) Hamiltonian are obtained by the inverse Dyson map:
\begin{equation}\label{eq:psi_from_phi}
\Psi_{n}(z) = S^{-1}\,\varphi_n(z)\,|\chi\rangle,
\end{equation}
where we take \(|\chi\rangle\) to be a \(\sigma_z\)-eigenvector,
\begin{equation}
\sigma_z|\chi_s\rangle = s\,|\chi_s\rangle,\qquad s=\pm1.
\end{equation}
Since \(S^{-1}=\exp\!\big(\mp\tfrac{m\alpha}{\hbar}z\,\sigma_z\big)\), the pulled-back spinor reads
\begin{equation}\label{eq:psi_n_unreg}
\Psi_{n,s}(z) = \mathcal{N}_n\, e^{\,s\frac{m\alpha}{\hbar}z}\,\mathrm{Ai}(\lambda z - a_n)\;|\chi_s\rangle.
\end{equation}

To test normalizability consider the large-\(z\) asymptotic of the Airy function:
\begin{equation}\label{eq:airy_asymp}
\mathrm{Ai}(\zeta)\sim\frac{1}{2\sqrt{\pi}}\,\zeta^{-1/4}\exp\!\Big(-\tfrac{2}{3}\zeta^{3/2}\Big),\qquad \zeta\to+\infty.
\end{equation}
Setting \(\zeta=\lambda z-a_n\sim\lambda z\) gives the leading tail of the density
\begin{equation}\label{eq:tail}
|\Psi_{n,s}(z)|^2 \propto \exp\!\Big(2s\frac{m\alpha}{\hbar}z \;-\; \tfrac{4}{3}(\lambda z)^{3/2}\Big),
\end{equation}
so the exponent contains a linear term (from the Dyson factor) and a superlinear negative term (from the Airy tail). Because \((\lambda z)^{3/2}\) grows faster than \(z\) as \(z\to\infty\), the negative superlinear term dominates for all finite \(\alpha\), and thus
\begin{equation}
\int_0^\infty |\Psi_{n,s}(z)|^2\,dz < \infty\qquad\text{for both }s=\pm1.
\end{equation}

\section{Proof of Reversibility}
\label{Appendix:ProofReversibility}
One can proof that using \(u:=S\psi\) and \(v:=S\phi\). Using \(\Theta=S^{-1}\mathcal T S\) and \(\eta=S^\dagger S\),
\begin{align*}
\big\langle \Theta \phi \big|_\eta \, \Theta \psi \big\rangle
&= \big\langle S^{-1}\mathcal T S \phi \;\big|\; \eta \;\big|\; S^{-1}\mathcal T S \psi \big\rangle\\[4pt]
&= \big\langle S^{-1}\mathcal T v \;\big|\; S^\dagger S \;\big|\; S^{-1}\mathcal T u \big\rangle.
\end{align*}
Now use \((S^{-1})^\dagger=(S^\dagger)^{-1}\). Hence
\[
\big\langle S^{-1} a \big| S^\dagger S \big| S^{-1} b \big\rangle
= \big\langle a \big| b \big\rangle
\qquad\text{for any }a,b,
\]
so with \(a=\mathcal T v,\; b=\mathcal T u\) we get
\[
\big\langle \Theta \phi \big|_\eta \, \Theta \psi \big\rangle
= \big\langle \mathcal T v \big| \mathcal T u \big\rangle.
\]
For an antiunitary \(\mathcal T\) the inner product reverses order:
\[
\big\langle \mathcal T v \big| \mathcal T u \big\rangle = \big\langle u \big| v \big\rangle.
\]
Therefore
\[
\big\langle \Theta \phi \big|_\eta \, \Theta \psi \big\rangle
= \big\langle u \big| v \big\rangle
= \big\langle S\psi \big| S\phi \big\rangle
= \big\langle \psi \big| S^\dagger S \big| \phi \big\rangle
= \big\langle \psi \big|_\eta \phi \big\rangle.
\]
Taking complex conjugates and using \((\langle\psi|_\eta\phi\rangle)^*=\langle\phi|_\eta\psi\rangle\) yields
\[
\big\langle \Theta \phi \big|_\eta \, \Theta \psi \big\rangle^{*}
= \langle\phi|_\eta \psi\rangle,
\]
which proves the claim.

\end{document}